\documentclass[apjl]{emulateapj}
\usepackage{amssymb}
\usepackage{natbib}
\usepackage{amsmath}
\usepackage{subfigure}
\usepackage{footnote}
\usepackage{color}
\usepackage{mdwlist}
\usepackage{enumerate}
\usepackage{graphicx}
\newcommand{\feviii}{{\ion{Fe}{8}}}
\newcommand{\fexiv}{{\ion{Fe}{14}}}
\newcommand{\fexii}{{\ion{Fe}{12}}}
\newcommand{\fexiii}{{\ion{Fe}{13}}}
\newcommand{\siiv}{{\ion{Si}{4}}}
\newcommand{\sivii}{{\ion{Si}{7}}}
\newcommand{\ov}{{\ion{O}{5}}}
\newcommand{\oiv}{{\ion{O}{4}}}
\newcommand{\oi}{{\ion{O}{1}}}
\newcommand{\ovi}{{\ion{O}{6}}}
\newcommand{\nixvi}{{\ion{Ni}{16}}}
\newcommand{\cii}{{\ion{C}{2}}}
\newcommand{\mgii}{{\ion{Mg}{2}}}

\newcommand{\heii}{{\ion{He}{2}}}
\newcommand{\fexi}{{\ion{Fe}{11}}}
\newcommand{\caxvii}{{\ion{Ca}{17}}}
\newcommand{\feix}{{\ion{Fe}{9}}}
\def\arcsec{$^{\prime\prime}$}

\newcommand{\longacknowledgement}{This work is supported by NASA under contract NNG09FA40C ({\it IRIS}).
PT was supported by contracts 8100002705 and SP02H1701R from Lockheed-Martin to SAO, and NASA contract NNM07AB07C to SAO. The work of DHB was performed under contract to the Naval Research Laboratory and was funded by the NASA \textit{Hinode} program.  
IRIS is a NASA small explorer mission developed and operated by LMSAL with mission operations executed at NASA Ames Research center and major contributions to downlink communications funded by ESA and the Norwegian Space Centre. \textit{Hinode} is a Japanese mission developed and launched by ISAS/JAXA, with NAOJ as domestic partner and NASA and STFC (UK) as international partners. It is operated by these agencies in co-operation with ESA and NSC (Norway).
CHIANTI is a collaborative project involving researchers at the universities of Cambridge (UK), George Mason and Michigan (USA). }
\begin{document}

\title{\textit{IRIS} observations of the low-atmosphere counterparts of active region outflows}
\author{Vanessa Polito}
\affil{Bay Area Environmental Research Institute, NASA Research Park,  Moffett Field, CA 94035, USA}
\affil{Lockheed Martin Solar \& Astrophysics Lab, Org. A021S, Bldg. 252, 3251 Hanover Street Palo Alto, CA 94304, USA}
\author{Bart De Pontieu}
\affil{Lockheed Martin Solar \& Astrophysics Lab, Org. A021S, Bldg. 252, 3251 Hanover Street Palo Alto, CA 94304, USA}
\affil{Rosseland Centre for Solar Physics and Institute of Theoretical Astrophysics,  University of Oslo, P.O. Box 1029
Blindern, NO-0315 Oslo, Norway}
\affil{Institute of Theoretical Astrophysics, University of Oslo,
P.O. Box 1029 Blindern, NO0315, Oslo, Norway}
\author{Paola Testa}
\affil{Harvard-Smithsonian Center for Astrophysics, 60 Garden Street, Cambridge MA 01238, USA}
\author{David H. Brooks}
\affil{College of Science, George Mason University, 4400 University Drive, Fairfax, VA 22030}
\author{Viggo Hansteen}
\affil{Bay Area Environmental Research Institute, NASA Research Park,  Moffett Field, CA 94035-0001, USA}
\affil{Lockheed Martin Solar \& Astrophysics Lab, Org. A021S, Bldg. 252, 3251 Hanover Street Palo Alto, CA 94304, USA}
\affil{Rosseland Centre for Solar Physics and Institute of Theoretical Astrophysics,  University of Oslo, P.O. Box 1029
Blindern, NO-0315 Oslo, Norway}
\affil{Institute of Theoretical Astrophysics, University of Oslo,
P.O. Box 1029 Blindern, NO0315, Oslo, Norway}
\begin{abstract}
    Active region (AR) outflows have been studied in detail since the launch of \textit{Hinode}/EIS and are believed to provide a possible source of mass and energy to the slow solar wind.   In this work, we investigate the lower atmospheric counterpart of AR outflows using observations from the \textit{Interface Region Imaging Spectrograph} (\textit{IRIS}). We find that the \textit{IRIS} \siiv, \cii\ and \mgii\ transition region (TR) and chromospheric lines exhibit different spectral features in the outflows as compared to neighboring regions at the footpoints (``moss") of hot AR loops. The average redshift of \siiv\ in the outflows region ($\approx$ 5.5~km s$^{-1}$) is  smaller than  typical moss ($\approx$ 12--13 km~s$^{-1}$) and quiet Sun ($\approx$ 7.5 km~s$^{-1}$) values, while the \cii~line is  blueshifted ($\approx$ -1.1--1.5 km~s$^{-1}$), in contrast to the moss where it is observed to be redshifted by about $\approx$ 2.5 km~s$^{-1}$. Further, we observe that the low atmosphere underneath the coronal outflows is highly structured, with the presence of blueshifts in \siiv\ and positive \mgii\ k2 asymmetries (which can be interpreted as signatures of chromospheric upflows) which are mostly not observed in the moss.  These observations show a clear correlation between the coronal  outflows and the chromosphere and TR underneath, which has not been shown before. Our work strongly suggests that these regions are not separate environments and should be treated together, and that current leading theories of AR outflows, such as the interchange reconnection model, need to take into account the dynamics of the low atmosphere.
    
\end{abstract}

\section{Introduction}

    Persistent upflows at the edges of active regions (AR outflows) are routinely observed by the \textit{Hinode} EUV Imaging Spectrometer \citep[EIS;][]{Culhane07} as blueshifts in the spectra of high-temperature coronal lines \citep[e.g.,][]{Harra08,DelZanna08,Doschek08, Baker09, Brooks11, Hinodereview12}.  Stronger blueshifts are typically observed in hotter lines \citep[T~$\approx$~1--2~MK, e.g.][]{DelZanna08}, which often exhibit blue asymmetries that also increase as a function of temperature \citep[e.g.][]{Bryans10}. We note that blueshifts at the edge of ARs were also observed before the launch of Hinode \citep{Marsch04}.
    %In contrast, redshifts in cooler lines  (T~$\approx$~0.15--0.8~MK) are  observed in the fan loops at the periphery of ARs \citep[e.g.][]{Young12}, although these structures do not seem to be related to the outflows \citep[e.g.][]{Warren11}. 
   \begin{figure*}
\centering
\includegraphics[width=\textwidth]{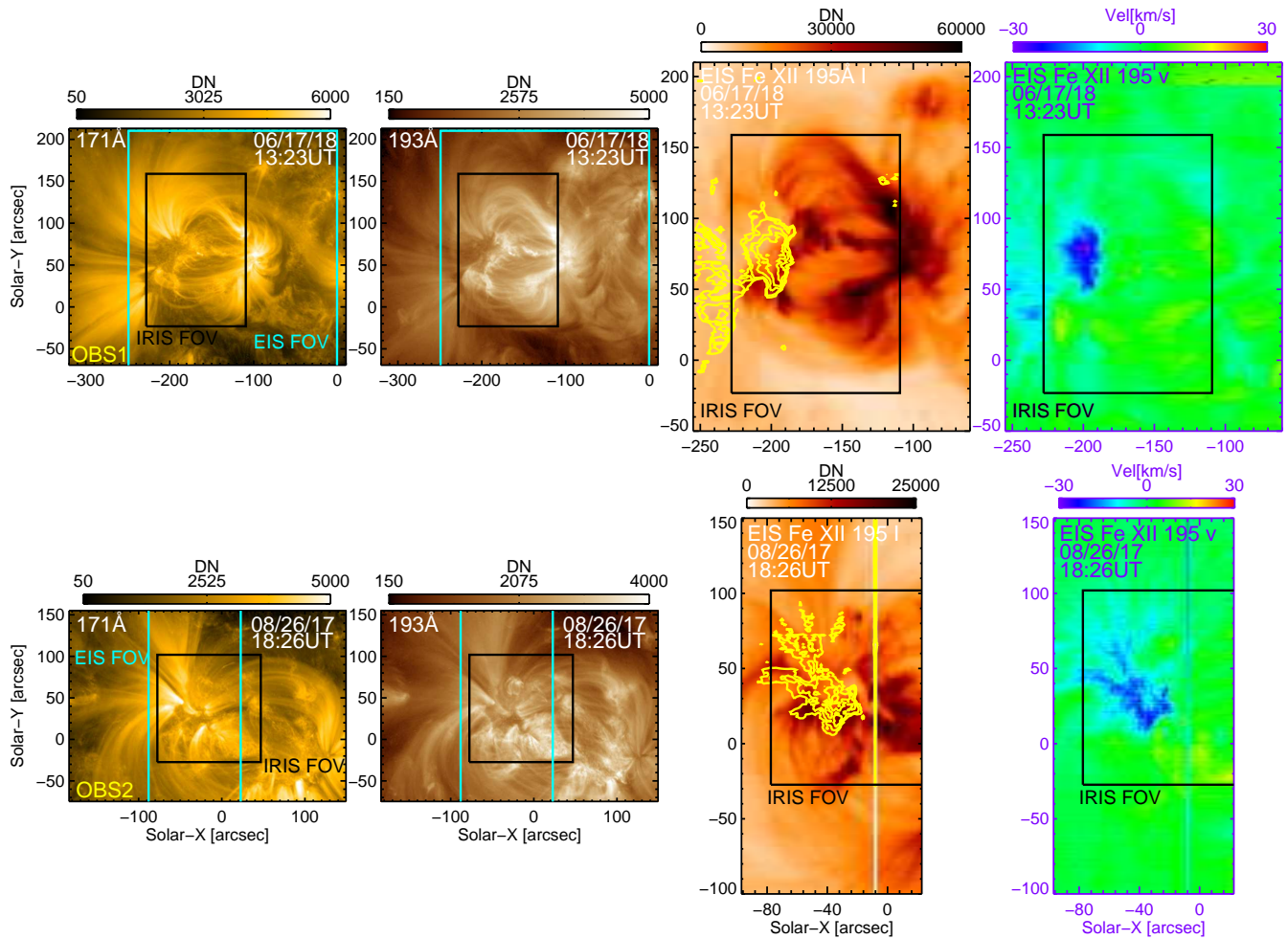}
\caption{From left to right: AIA 171~\AA\ and 193~\AA\ images of the two AR observations under study, with the EIS and \textit{IRIS} FOVs overlaid; intensity and Doppler velocity maps of the EIS \fexii~195~\AA\ line for OBS1 (top) and OBS2 (bottom). }
\label{Fig:context}
\end{figure*}

    The AR outflows have often been invoked as a possible source of slow solar wind \citep[e.g.][]{Sakao07}. EIS observations, combined with in-situ measurements, have shown that the outflows have a slow-wind plasma composition, 
    in agreement with this hypothesis \citep[e.g.][]{Brooks11, Brooks12, Brooks15}. Recent high-resolution observations from Hi-C 2.1 \citep{Rachmeler19} have also %demonstrated the existence of 
    suggested the existence of photospheric and coronal components to the outflows, 
    %two components to the outflows, 
    which may explain the variable composition of the slow wind \citep{Brooks20}.
    
    Nevertheless, a definitive model to explain how the outflows are driven and connected to the slow wind is still lacking. 
    %outflows could be either flowing into very long and low density loops or travelling to the SW along open field lines \citep[e.g.][]{Culhane07, Doschek08,Boutry12}. 
    %? See also discussion. 
    A popular scenario assumes that interchange reconnection between closed dense AR loops and neighbouring open (or distantly connected) large-scale low-density loops may act as a mechanism for driving the outflows \citep[e.g.][]{Baker09,DelZanna11}. 
    
    One issue with this picture is that the outflows are not always observed in the vicinity of open field lines \citep[e.g.][]{Culhane14}, and \cite{Boutry12} estimated that a significant fraction of the outflows propagate into long loops connected to distant ARs. A solution might be provided by a ``two-step" reconnection process \citep{Culhane14,Mandrini15}, where closed loop plasma first travels along long loops and then, in a second step, is released into the open field lines via interchange reconnection.

   Alternative scenarios suggest that the outflows are linked to chromospheric jets and spicules, which are heated while propagating into the corona \citep{McIntosh09,DePontieu09,DePontieu17}.

   The height at which the outflows are driven, whether in the corona or the lower atmosphere, is still under debate. 
    \begin{figure*}
\centering
\includegraphics[width=\textwidth]{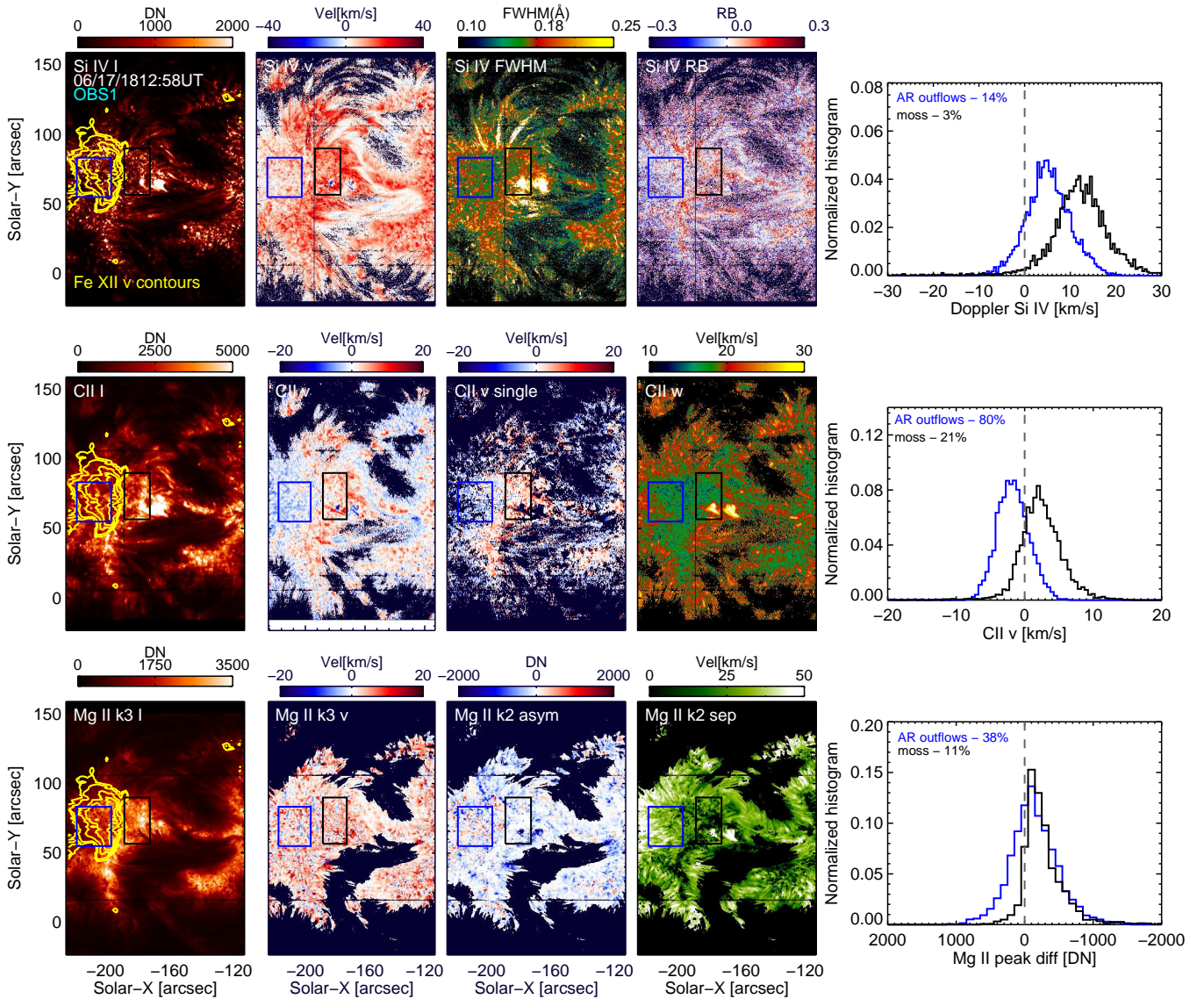}
\caption{Top rows (from left to right): total intensity, Doppler shifts, FWHM  and RB asymmetry maps, and histograms of Doppler shifts of \siiv~in the outflow and moss boxed regions. Middle rows: first and second moments, second moment for single peaked profiles only and third moment of the \cii~line, and histograms of the second moment in the boxed regions. Bottom rows: \mgii~k3 intensity, velocity, k2 peak asymmetry and peak separation, and histograms of the k2 peak asymmetry in the boxed regions. Contours of the EIS \fexii~outflows are also overlaid in yellow. The histograms are normalised by the number of pixels in each box. The percentage of pixels with negative velocity (for \siiv~and \cii) or positive counts (for \mgii~asymmetry) for both moss and outflow regions is also reported in each panel.}
\label{Fig:fig2}
\end{figure*}
 \cite{Vanninathan15} suggested that there was not enough chromospheric jet-like activity or spectral asymmetries in H$\alpha$ to account for a connection with outflows. On the other hand, \cite{He10} found  intermittent upflows in \textit{Hinode}/XRT images, associated with chromospheric jets observed by \textit{Hinode}/SOT and blueshifts in the EIS \heii~line, suggesting a possible chromospheric origin of the outflows.
 
These previous studies showcase the importance of observing the evolution of the outflows throughout  different layers of the solar atmosphere. In this work, we combine EIS observations of coronal outflows with recent spectroscopic observations from \textit{IRIS} \citep{DePontieu14} at high spatial (0.33--0.4\arcsec) and spectral ($\approx$~1.5 km~s$^{-1}$) resolution. We focus on studying emission from \siiv~1393.75~\AA~(TR), \cii~1335.71~\AA~(upper chromosphere to TR) and \mgii~k~2796.35~\AA~(mid to upper chromosphere)  \textit{IRIS} lines.

\section{Observations}
\label{Sect:obs}
We select two ARs that were simultaneously observed by \textit{IRIS} and EIS: AR 12713 on 2018-06-17 (OBS1) and AR 12672 on 2017-08-26 (OBS2, Figure~\ref{Fig:context}), using large rasters that covered both the moss and outflows at the edge of the ARs. We selected ARs %were 
close to disk center, to minimize possible projection effects  \citep{Baker17,Gosh19}. The \textit{IRIS} datasets were both dense (slit width = ~0.33\arcsec) 320-step rasters, with exposure times t$_{exp}$ of 8s and 15s respectively. The EIS OBS1 study was a 87-step raster taken with a raster step of 3\arcsec, slit width of 2\arcsec~and t$_{exp}$~=~40s, while OBS2 was a dense 60-step raster with a raster step of 2\arcsec\ and slit width of 1\arcsec\ and t$_{exp}$~=~45s.

Figure~\ref{Fig:context} provides a context view of the observations. The two leftmost panels show the ARs under study as observed by AIA in the 171~\AA\ and 193~\AA\ filters, which in ARs are typically dominated by emission from \feix~(logT(K)~$\approx$~5.85) and \fexii~(logT(K)~$\approx$~6.2), respectively \citep{O'Dwyer10,Martinez11}. The field of view (FOV) of the \textit{IRIS} and EIS rasters are also overlaid. The third and fourth panels show the intensity and Doppler shifts of the EIS \fexii~195.12~\AA~line, which were obtained by performing a single Gaussian fit in each pixel. The \fexii\ blueshifts highlight the outflow region, while the bright emission at the footpoints of the hot AR core loops, visible in both the AIA 193~\AA~and EIS \fexii~images, characterizes the moss.

\begin{figure*}
\centering
\includegraphics[width=\textwidth]{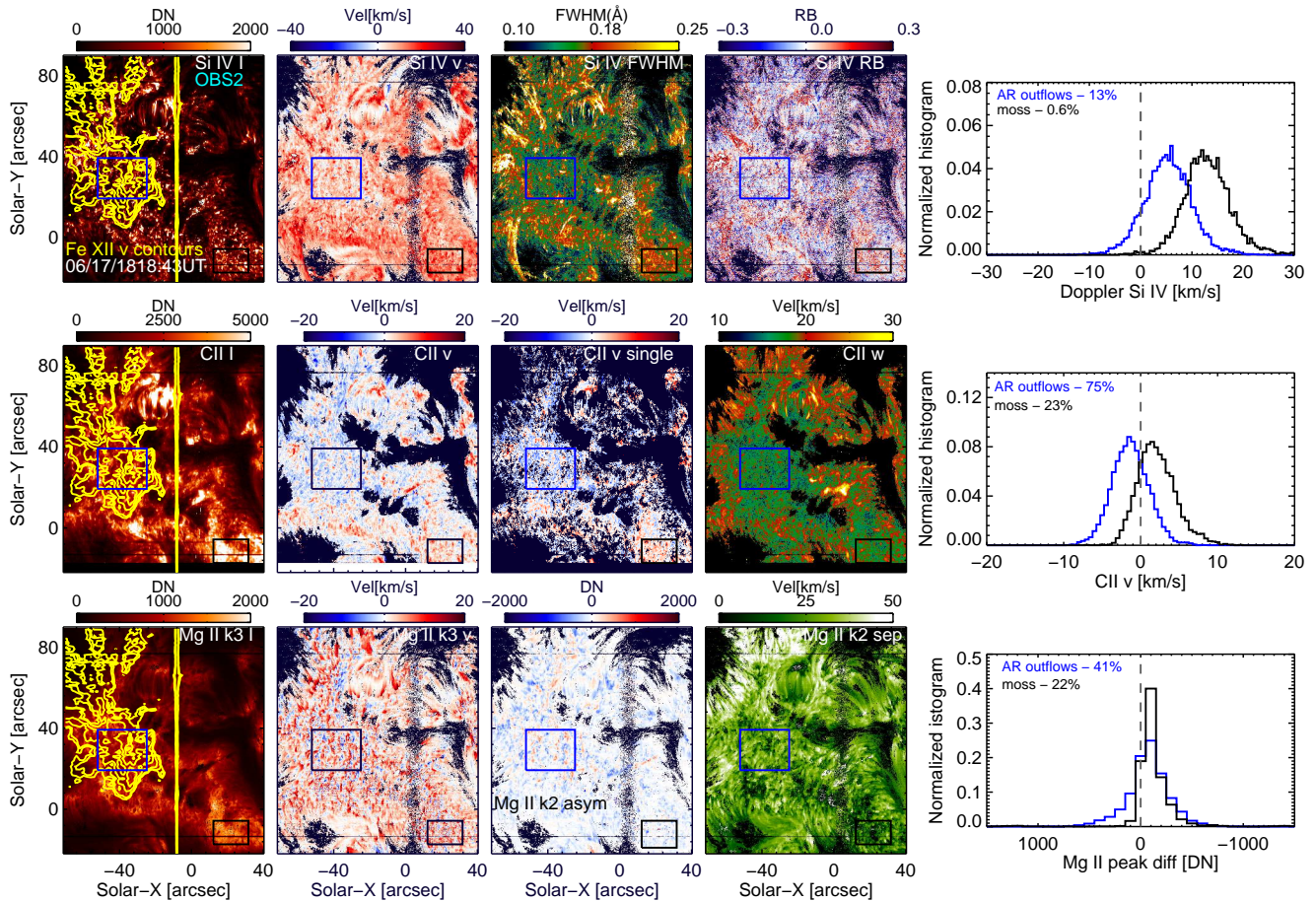}
\caption{Same as Fig.~\ref{Fig:fig2} for OBS2.}
\label{Fig:fig3}
\end{figure*}
We used \textit{IRIS} level 2 data that has been corrected for geometrical, flat-field and dark current effects and for the orbital variation of the wavelength array. Further, we use the \oi\ photospheric line to check the absolute wavelength calibration. After summing the \oi\ spectra in a quiet region outside the AR, we estimated that the average velocity shift from the at-rest position of 1355.598~\AA~was around 0.5~km~s$^{-1}$. 

The EIS data are affected by a series of issues 
which are
described at length in the instrument documentation\footnote{http://solarb.mssl.ucl.ac.uk/eiswiki/Wiki.jsp?page=EISAnalysisGuide}, and that are carefully taken into account in our analysis. Given the lack of reference photospheric lines to perform an absolute wavelength calibration, the uncertainty in the EIS Doppler shift velocities is usually estimated to be at best $\approx$~5 km~s$^{-1}$ \citep[e.g.][]{Young12}.

To co-align the IRIS and EIS spectroscopic observations, we first compared the SJI 1330~\AA~ \citep[logT(K)~$\approx$~4.3,][]{DePontieu14} and AIA 1700~\AA~ \citep[logT(K)~$\approx$~4.2,][]{Simoes19} images, which are dominated by plasma formed at similar temperatures. The EIS images formed in the \heii~256~\AA~line (logT(K)~$\approx$~4.7) were also co-aligned to AIA 304~\AA~images, which are dominated by emission from \heii~\citep{O'Dwyer10,Martinez11}. To co-align all the images we used standard cross-correlation routines available in \textit{SolarSoft}, and also verified the alignment by eye. Using these different methods we estimate a co-alignment uncertainty between \textit{IRIS} and EIS of $\approx$~4\arcsec, which is reasonable considering that the full-width half maximum (FWHM)  of the EIS point-spread-function is $\approx$~3\arcsec.

 \subsection{IRIS observations of outflows}
In this Section, we focus on observing the spectral features of the \textit{IRIS} lines in the region underneath
the coronal outflows. Figures~\ref{Fig:fig2} and \ref{Fig:fig3}  provide an overview of the \textit{IRIS} observations in \siiv~(top), \cii~(middle) and \mgii~(bottom) for OBS1 and 2 respectively.  The first 
row %panel 
shows total intensity, Doppler shift, and FWHM of the \siiv~line, assuming a single Gaussian fit, as well as a red-blue (RB) asymmetry map, which was calculated using the algorithm described in \cite{Tian11} with a $\pm$~40~km~s$^{-1}$ velocity interval and a 5~km~s$^{-1}$ bin. Zero-th (intensity), first (velocity) and second (standard deviation, expressed in km~s$^{-1}$) moments of the \cii\ line are shown in the second row. A map of first moments, only for profiles which are identified as single peaked, is also shown in the third panel of the second row. Being optically thick, the \cii\ line has a complex formation and can exhibit two or more peaks. We identified the single peaked profiles using the method described in \cite{Rathore15}. The bottom rows of Figs.~\ref{Fig:fig2} and \ref{Fig:fig3} show the \mgii\ 2796~\AA\ line k3 (line core) intensity and velocity, the k2 peak asymmetry (difference between the intensity of the red and blue k2 peaks) and peak separation (difference between the wavelengths of the red and blue k2 peaks, expressed as velocity). The k3 Doppler shifts provide velocity diagnostics in the upper chromosphere, where the line core is formed, while the k2 asymmetry and separation have been shown to provide good diagnostics of flows and velocity gradients \citep{Leenarts13}. In particular, a positive/negative k2 asymmetry is typically indicative of up/downflows (because it is caused by increased absorption in the blue/red wavelengths).  The contours of the \fexii\ outflows are also overlaid in yellow. Finally, each row includes a normalised histogram of Doppler velocities (for \siiv\ and \cii, with bin of 5 km~$s^{-1}$) or k2 asymmetries (for \mgii, with a bin of 100 DN) inside the two boxes overlaid in the respective variable maps (black and blue boxes for moss and outflow regions respectively). The percentage of pixels with negative velocity (for \siiv\ and \cii) or positive counts (for \mgii) are also reported in each panel for both moss and outflow regions.
Further, we  calculated the percentage of pixels taking into account the estimated error on our velocities (0.5 km~s$^{-1}$) by setting the threshold velocity to 0.5 and -0.5 km~s$^{-1}$. The  percentages of \siiv~upflows and their associated range of values taking into account this velocity uncertainty are: 14.4$\%$ (12--17$\%$) in the outflows and 3.3$\%$ (3.1--3.7$\%$) in the moss for OBS1, and 12.9$\%$~(11--15$\%$) 
in the outflows and 0.6$\%$ (0.5--0.7$\%$) in the moss for OBS2. These values suggest that the fraction of pixels showing \siiv~upflows in the outflow region is significantly higher (at least 4 times higher in OBS1) than that in the moss.

Figures~\ref{Fig:fig2} and \ref{Fig:fig3} show some clear features that distinguish the outflow from the moss region: (1) the average distribution of \siiv\ Doppler shift is shifted towards the blue in the outflow regions, with an average value of $\mu_{out}$~$\approx$~5.4--5.6~km~s$^{-1}$ and standard deviation $\sigma_{out}$~$\approx$~4.8--4.6~km~s$^{-1}$ for OBS1 and 2 (compared to the moss values of  $\mu_{moss}$~$\approx$~12.6--13.3~km~s$^{-1}$ and $\sigma_{moss}$~$\approx$~7.4--4.6~km~s$^{-1}$ respectively); (2) \siiv~upflows up to $\approx$~-10 km~s$^{-1}$ are observed in the outflow region but not in the moss, where the line is consistently redshifted; (3) a similar trend is visible in \cii, with the difference that the velocity distributions in outflows and moss regions are both shifted towards the blue compared to those of \siiv. The values for \cii~are: $\mu_{out}$~$\approx$~-1.4 and -1.1~km~s$^{-1}$,  $\sigma_{out}$~$\approx$~2.3 and 2.4 ~km~s$^{-1}$, in contrast to $\mu_{moss}$~$\approx$~2.6 and 2.4~km~s$^{-1}$, $\sigma_{moss}$~$\approx$~3.5 and 2.5~km~s$^{-1}$ for OBS1 and OBS2 respectively; (4) we observe a higher concentration of \mgii\ profiles with positive k2 asymmetry (i.e. upflows) in the outflows than in the moss region. 

The \oi\ 1355.598~\AA\ line also appears to be more blueshifted in the outflows compared to the moss. However, since the line is faint, the Doppler shift measurements can be uncertain. No peculiar features are observed in the \mgii\ triplet lines in the outflow regions.

\subsection{Combining IRIS and EIS observations for OBS1}

\begin{figure*}
\centering
\includegraphics[width=\textwidth]{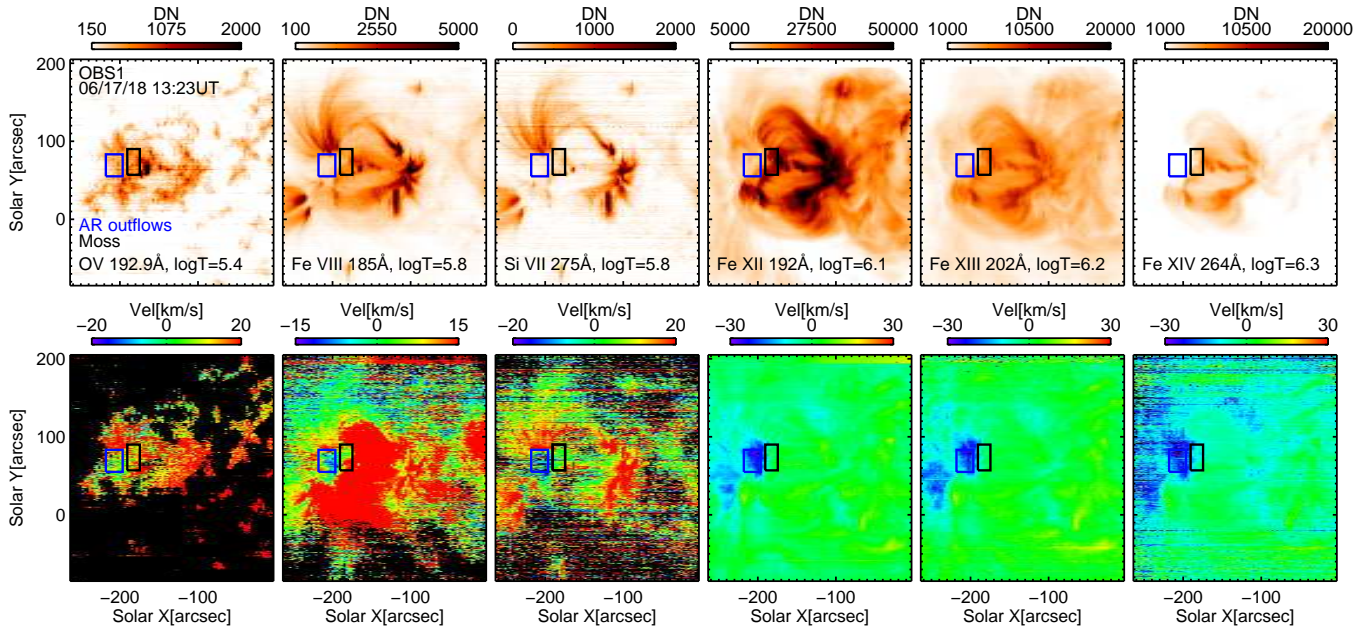}
\caption{Intensity (top) and Doppler shift (bottom) maps of EIS lines formed at different temperatures for OBS1. The outflows and moss boxed regions (same as those in Fig.~\ref{Fig:fig2}) are overlaid on each panel.  }
\label{Fig:fig4}
\end{figure*}

Figure~\ref{Fig:fig4} shows an overview of the EIS lines (OBS1) formed over a broad range of temperatures. 
The \ov\ 192.9~\AA\ line is close in wavelength to the hot \fexi\ 192.83~\AA\ and \caxvii\ 192.82~\AA\ lines \citep{Young07}, but can be usually reliably resolved using a multi-Gaussian fit. We note that the unblended \ov\ line at 248.46~\AA\ was not included in this observation program, while other oxygen TR lines such as \oiv\ and \ovi\ are too faint. The \feviii\ 185.21~\AA\ line is also blended with the hotter \nixvi\ 185.23~\AA\ line which is mostly visible in the AR cores \citep{Young07}. The \feviii\ line is formed at the same temperature as the \sivii\ 275.75~\AA\ line, which is largely free of blends and also shown in Fig.~\ref{Fig:fig4}. Both \feviii\ and \sivii\ show some faint blueshifted emission in the outflow region.

The strong \fexii~195.12~\AA~line is self-blended with a \fexii~transition at 195.18~\AA~that can become more significant at high densities \citep[$>$~10$^{10}$~cm$^{-3}$,][]{DelZanna05}. The weaker \fexii~192.39~\AA~and the \fexiii~202.04~\AA~(logT(K)~$\approx$~6.25) lines are also included in this study and they are both mostly free of blends \citep{Young07,DelZanna11b}.

\begin{figure}
\centering
\includegraphics[width=0.5\textwidth]{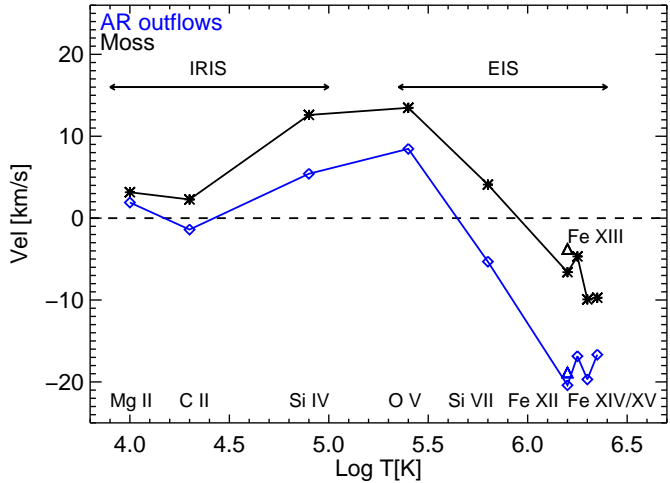}
\caption{Velocity vs temperature in outflows and moss for different EIS and \textit{IRIS} lines. }
\label{Fig:fig5}
\end{figure}

Figure~\ref{Fig:fig5} summarizes the average velocities observed by EIS and \textit{IRIS} in the moss and outflows regions of OBS1. These values have been obtained by averaging the velocity inside the boxes highlighted in Fig.~\ref{Fig:fig2} and Fig.~\ref{Fig:fig4}. For \fexii, we take the Doppler shifts of the 192.39~\AA~line, but also include the values obtained for the 195.12~\AA~line (triangles) for reference. For the \mgii\ line we used the k3 Doppler shift. 

Figure~\ref{Fig:fig5} shows a similar trend of increasing blueshifts with temperature for the EIS hotter lines (logT(K)~$\approx$~5.8--6.3, e.g. \sivii--\fexiv), 

 as that reported in Figure 3 of \cite{DelZanna08}. The cooler EIS and \textit{IRIS} lines (logT(K)~$\approx$~4--5.4) show that the trend is almost inverted in the lower atmosphere, with the \ov, \siiv\ and \mgii\ lines being on average redshifted, while interestingly the \cii\ line is slightly blueshifted.  
 
For comparison, Figure~\ref{Fig:fig5} also shows the Doppler shift trend for the moss region. The blueshifts  are consistently smaller while the redshifts are larger in the moss as compared to the outflow region, although the trend of Doppler shifts vs temperature appears to be similar in the two regions.

It is important to note that Figs.~\ref{Fig:fig2} and \ref{Fig:fig3} reveal a more complicated picture than that suggested by the averaged trend of Fig.~\ref{Fig:fig5}, with the presence of small-scale TR and chromospheric upflows in the high-resolution \textit{IRIS} data.

\section{Discussion and conclusions}
\label{Sect:discussion}

 \textit{IRIS} TR and chromospheric observations offer a  unique view of the lower atmospheric counterpart of the long-studied AR outflows at very high spatial and spectral resolution.  
We analysed two observations of AR outflows with \textit{IRIS} and EIS and found that: 

\begin{itemize}

\item \textit{IRIS} reveals that the low atmosphere underneath the coronal outflows appears to be highly structured, with a percentage of $\approx$~13$\%$ pixels showing small scale upflows in \siiv\ (up to $\approx$~-10 km~s$^{-1}$), as compared to 3~$\%$ or less in the moss.

    \item The \siiv\ TR line is also on average less redshifted in the outflow region ($\approx$~5.5 km~s$^{-1}$) than in the moss ($\approx$~13~km~s$^{-1}$). The average \siiv\ redshift in the outflows is also lower than typical QS values (7.5 km~s$^{-1}$ from an \textit{IRIS} sample observation, in agreement with measurements from previous instruments, e.g. \citealt{Doschek76,Brekke97}). 
    In contrast with the moss region, where the profiles show some red asymmetry, the \siiv~spectra show no asymmetry or a modest blue asymmetry in the outflow region. 
   \item The \mgii\ k2 asymmetry maps show more profiles with a higher red peak in the outflow region than in the moss. This can be interpreted as evidence of upflows \citep[i.e., more absorption in the blue wing,][]{Leenarts13,SainzDalda19}. The average k3 velocities in the outflow region are also smaller ($\approx$~2~km~s$^{-1}$) compared to the moss ($\approx$~3~km~s$^{-1}$) .
 
   \item While being redshifted in the moss ($\approx$~2.5 km~s$^{-1}$), the \cii\ line is on average sligthly blueshifted ($\approx$~-1.1--1.5~km~s$^{-1}$) in the outflow region. As a comparison, 
   \cite{Rathore15} previously reported median values of \cii\ Doppler shifts (calculated assuming Gaussian fits) of $\approx$~0~km~s$^{-1}$ in  the QS network and plage, and $\approx$-2.3~km~s$^{-1}$ in the QS internetwork.

\end{itemize}
Our findings show that there is a clear correlation between the chromosphere and TR observed by \textit{IRIS} and the coronal AR outflows observed by EIS.
Such correlation has never been shown before and is surprising given previous work which found no connection between chromospheric activity and coronal outflows using H$\alpha$ data from IBIS \citep{Vanninathan15}. However, it is not clear why such a connection should make itself visible through the asymmetry of an optically thick line with a complex formation mechanism like H$\alpha$. Our results suggest that the restricted focus on blue-ward asymmetries in the H$\alpha$ line misses the actual association between chromospheric and TR signals and coronal outflows.

There are no models that can currently explain the regional differences of chromospheric and TR properties between outflow regions and moss or QS. We cannot determine from these preliminary studies whether this association occurs because the mechanism that drives outflows occurs at these low heights, or because the outflow regions are actually driven in the larger coronal volume but have a different coronal thermodynamic environment that indirectly affects the lower-lying regions, due for example to the fact that AR outflows may be connected to open field regions, as it is commonly assumed. Comparisons between the closed field regions in QS and open field regions in coronal holes (CHs) have reported subtle differences with CH network showing somewhat less redshifted profiles in low TR spectral lines \citep{Mcintosh11}, similarly to what we observe here. However, from a preliminary analysis, other properties such as the \siiv~line widths and \mgii~spectral properties seem to behave differently in CHs and  AR outflows.

Previous work has demonstrated that the dynamics of the upper chromosphere and low TR in plage are strongly affected by processes that are driven from the lower atmosphere. For instance, magneto-acoustic shocks are ubiquitous in plage \citep{Hansteen06} and have a major impact on the properties of the \mgii~ h\&k,  \cii~and \siiv~lines \citep[e.g.,][]{DePontieu15}. Similarly, the disk counterparts of type II spicules lead to sudden blueshifted excursions in chromospheric lines and broadened linear features in \siiv~ \citep{Tian14,Rouppe15, DePontieu17}. Perhaps the modification of these phenomena in the open fields of outflow regions plays a role in explaining our results? For example, it is well known that spicules are taller and have different dynamical properties in open fields \citep{Beckers68}. There are also suggestions from numerical models that the magnetic topology of a region plays an important role and that TR emission and flow patterns are significantly different in short, low-lying loops (that are not connected to the corona), from those in longer, higher coronal loops, with the height of heating determining the strength of TR redshifts and/or blueshifts \citep{Guerreiro2013,Hansteen2014}. Perhaps the observed differences in TR properties are related to a different proportion of low-lying to higher loops or that heating occurs at a different height in AR outflow sites compared to moss regions (or CHs)? 

More extensive statistical studies and numerical modeling are clearly required to settle these issues.  The exact impact of the open fields on the upper chromosphere and low TR thus remains uncertain. 

Nevertheless, our work strongly suggests that the low  atmosphere and the corona should not be treated as separate environments, but rather that the processes occurring throughout different layers of the solar atmosphere are likely connected, and should be addressed together by a global model. This is also supported by the fact that the leading theories to explain the first ionization potential (FIP) effect  are based on the idea that the FIP fractionation most likely takes place in the upper chromosphere \citep[e.g.][]{Laming15,Laming19}.

So how do our \textit{IRIS} observations fit within the context of the most recent models of outflows? 
These are based on interchange reconnection between closed and open field lines. Modelling by \citet{Bradshaw11} (in 1D) and \citet{Harra12} (in 3D) has demonstrated that coronal outflows could be explained as pressure-driven flows following reconnection between closed and open field lines. However, both papers mostly focused on the synthesis of hotter emission observed by EIS. Our work suggests that future outflow models need to include a proper treatment of the chromosphere and TR (i.e., including optically thick radiative losses, convective motions, magneto-acoustic shocks, and spicules) to explain both EIS and \textit{IRIS} observations. Such models may be able to elucidate whether waves and heating associated with spicules \citep{DePontieu17} plays any role in outflow regions, possibly related to differences in the underlying photospheric magnetic fields \citep{Samanta19}. 

Our work demonstrates that to understand the origin of AR outflows it is crucial to follow the plasma dynamics through different layers of the atmosphere.  Our results raise several unsolved questions about the nature of outflow regions and provide a challenge to previous and future models on outflows. Coordinated observations between \textit{IRIS} and high-sensitivity radio observatories, such as the Jansky Very Large Array, as well as the recent Parker Solar Probe and Solar Orbiter missions, will  be key to obtain a more complete picture of the outflow regions and their connection to the solar wind.

\acknowledgements{\longacknowledgement}

\bibliographystyle{aasjournal}

\bibliography{outflows}

\end{document}